\documentclass[twocolumn,prl,showpacs,nofootinbib,superscriptaddress]{revtex4}
\usepackage{latexsym}
\usepackage{dcolumn}
\usepackage{epsfig}

\newcommand{\ba}{\begin{eqnarray}}

\newcommand{\ea}{\end{eqnarray}}
\newcommand{\be}{\begin{equation}}
\newcommand{\ee}{\end{equation}}

\newcommand{\eq}[1]{Eq.\,(\ref{#1})}

\newcommand{\sigtot}{\sigma_{\rm tot}}

\def\bea{\begin{eqnarray}} 
\def\eea{\end{eqnarray}}

\begin{document}

\title{Commentary on ``Total Hadronic Cross Section Data and the Froissart-Martin Bound '', by Fagundes, Menon and  Silva}

\author{Martin~M.~Block}
\affiliation{Department of Physics and Astronomy, Northwestern University, 
Evanston, IL 60208}
\author{Francis Halzen}
\affiliation{Department of Physics, University of Wisconsin, Madison, WI 53706}

\begin{abstract}
This Commentary on the  paper, ``Total Hadronic Cross Section Data and the Froissart-Martin Bound '',  by  Fagundes, Menon and Silva (Braz. J. Phys., Vol. 42 (2012); arXiv:1112.4704) was invited by the Editors of the Brazilian Journal of Physics to appear directly after the above authors' printed version, in the same journal issue. We here challenge that paper's conclusions that the Froissart bound was violated.  We will show that this conclusion follows from a statistical  methodology that we question, and will present compelling supplementary evidence that the latest ultra-high energy experimental $pp$ cross section data are consistent with a $\ln^2 s$ behavior that satisfies the Froissart bound. 
\end{abstract}
\date{\today}
\pacs{ 12.38.Qk, 13.85Hd, 13.85Lg, 13.85Tp}

\maketitle

{\em Introduction.}---We have been invited by the Editors of the Brazilian Journal of Physics to write a  back-to-back Commentary,  to be published in the same printed journal,  on  ``Total Hadronic Cross Section Data and the Froissart-Martin Bound '',  by  D.~A.~Fagundes, M.~J. ~Menon and P.~V.~R.~G.~ Silva, Braz. J. Phys., Vol. 42 (2012); arXiv:1112.4704,  hereafter referred to as FMS \cite {FMS}. We wish to challenge their conclusion that the original  Froissart bound \cite{froissart} is violated, i.e.,  their statement that the total $pp$ ($\bar p p$) section  rises  more rapidly with energy  than $c\ln^2 s$, where $c$ is a constant and $s$ is the square of the hadron-hadron center of mass system (cms).

In essence, FMS conclude that when all total cross section data for $pp$ and $\bar p p$ between $5\le \sqrt s\le 1800$ GeV are used---a total of 163 datum points---the measured  total cross sections satisfactorily satisfy the relations 
\ba
\sigma_{LE}&=&a_1\left({s\over s_l}\right)^{-b_1}\pm a_2\left({s\over s_l}\right)^{-b_2},\label{sigmaLE}\\
\sigma_{HE}&=&\alpha +\beta\ln^\gamma {s\over s_h}, \quad \gamma =2, \label{sigmaHE}\\
\sigma_{\rm tot}&\equiv&\sigma_{LE}+ \sigma_{HE}\label{sigFMS},
\ea
where $a_1,a_2,b_1,b_2, \alpha,\beta,\gamma$ and $s_h$ are real constants and $s_l=1$ GeV$^2$ and the + sign is for  $\bar p p$ and the - sign is for $pp$ collisions, i.e., they find that the Froissart bound is satisfied,  but that is is violated  ($\gamma>2$)  when {\em one} additional point at 7 TeV---the Totem \cite{totem} total cross section measurement of $pp$, $\sigma_{\rm tot}=98.3\pm 0.2 ({\rm stat.})  \pm 2.8 ({\rm syst.}) $ mb---is considered. As examples,  see the $\gamma=2$ result from Table 1, Direct Fit of FMS, and $\gamma=2.104\pm 0.027$ from Table 2, V1 of FMS.  We will show that the statistical probability of the $\gamma=2$ solution  {\em with the inclusion} of the Totem result (Direct Fit model of Table 2, FMS) is {\em higher} than the statistical probability of the fit (Table 2, V1 model) that gave $\gamma>2$.  Thus, we find no statistical evidence for their claim. 

Further, since they only include {\em one} additional high energy point at 7 TeV in their analysis, an alternate (and in this case, perhaps more transparent) analysis method would have been to use the 163 $pp$ and $\bar p p$ cross section datum points in the energy interval $5\le \sqrt s\le 1800$ GeV to {\em predict} the 7 TeV total cross section. We will show that when we reanalyze the FMS results, the low energy fit for $\gamma=2$ after including the errors of prediction (due to statistical errors in the parameters) {\em satisfactorily predicts} the 7 TeV Totem total cross section measurement.

We will also comment on the necessity for the { simultaneous}  inclusion of the $\rho $ data, the ratio of the real to the imaginary portion of the total cross section, together with the cross section data, i,e., a {\em global fit} of $\rho$ and $\sigma_{\rm tot}$, a statement alluded to (but {\em not} carried out) in the Appendix of FMS. Since all but one of the parameters is {\em common} to both $\rho$ and $\sigma$, the simultaneous inclusion of the many $\rho$-values is important to an accurate fit, in order to minimize the (correlated) errors on the fitted parameters. We strongly disagree with the FMS comments 1 to 6 in their Section 2.2 concerning the $\rho$-values, in which they dismiss them as essentially useless, whereas in their Appendix,  they give a rather cumbersome evaluation using their Variant 3 model, to {\em separately evaluate} $\rho$, and {\em not globally evaluate} it with the total cross sections, as demanded by analyticity.

Recently, eight measurements of  $pp$ cross sections have been made at energies  higher than  the Tevatron energy of 1800 GeV. At the LHC cms energy of 2.76 TeV, ALICE \cite{alice}   has measured $\sigma_{\rm inel}$, at the LHC cms  energy of 7 TeV, ALICE \cite{alice}, ATLAS \cite{atlas}, CMS \cite{cms} and TOTEM \cite{totem} have measured $\sigma_{\rm inel}$; $\sigma_{\rm tot}$ has been measured by TOTEM \cite{totem}. Most recently,  the Pierre Auger Observatory has published $pp$ cross sections for $\sigma_{\rm inel}$ and $\sigma_{\rm tot}$ at 57 TeV \cite{auger}, using cosmic ray measurements of extended air showers to measure $\sigma_{\rm p-air}$. Most likely, this is effectively the highest energy reach that one will ever have experimentally.  

Finally, using these ultra-high energy measurements, we will summarize a very recent analysis of Block and Halzen \cite{blackdisk2} that shows that the proton asymptotically becomes a {\em black disk}, whose total cross section asymptotically varies with energy as a {\em saturated} Froissart bound, i.e., as $
\ln^2 s$, and whose asymptotic ratio of inelastic to total cross section is 1/2, that of the black disk.  This result, using {\em all available} high energy data, including inelastic cross sections,   clearly contradicts the conclusions of FMS. 

{\em Statistical Probabilities}---For a $\chi^2$ fit, the  goodness-of-fit criterion for a given model is normally taken as $P(\chi^2_0;\nu)$, the integral of the  probability distribution in $\chi^2$ for $\nu$ degrees of freedom, integrated from the observed minimum $\chi^2_0$ to infinity; it is $P(\chi^2_0;\nu)$ that allows us to statistically distinguish between models.   From Table 1 of FMS, we concentrate on the Direct Fit  (DF$_1$) and V1 models, which give $\chi^2_0=145.236,\ \nu=156,$ yielding $P_{\rm DF_1}=  0.721$, whereas for V1, we have  $\chi^2_0=145.235,\ \nu=155,$ yielding $P_{\rm V1}=  0.701$. Thus, even though FMS relaxed the input restriction that $\gamma=2$ and allowed it to be fit by the data, we get the somewhat strange result that FMS have a {\em better, somewhat more reliable} fit when they fix the value of $\gamma$ at 2, the Froissart bound limit, than when  they allow it to float, suggesting perhaps that the {\em true} minimum $\chi^2$ was not achieved in their minimization process.  In any event, FMS concluded that the value $\gamma=2$ was correct for the energy interval $5\le\sqrt s\le 1800$ GeV.

Next, we calculate the  goodness-of-fit probabilities for the Direct Fit and V1 models from Table 2, which now  includes {\em one additional point}, the Totem point at 7 TeV. For the Direct Fit  model (DF$_2$), we find  $\chi^2_0=146.01,\ \nu=157$, giving $P_{\rm DF_2}=  0.725$; this is effectively {\em identical} to the comparable value obtained ($P_{\rm DF_1}$=0.721) not using the Totem point. Thus, if $\gamma=2$ is satisfactory for the low energy data, it appears to be exactly  the same level of confidence when we include the Totem point.  Finally, we calculate from Table 2 the high energy version of the V1 model (V1$_2$), i.e., $\chi^2_0=145.86,\ \nu=156$, giving $P_{\rm V1_2}=  0.709$, a probability {\em smaller} than  that of the Direct Fit model (DF$_2$), i.e., $P_{\rm DF_2}= 0.725$.  Again, this result is somewhat strange, since the logarithmic power $\gamma$ was left adjustable in the V1$_2$ model.

Completely analogous results were found when  comparing the V4 and V5 models from Table 3 .  We obtain $P_{\rm V4_1}=0.616$ for fixed $\gamma=2$ and $P_{\rm V5_1}=0.595$ when FMS  let  $\gamma$ float for the low energy data---again, it is very strange that letting $\gamma$ be fit by the data gave a lower probability than fixing $\gamma$ to be 2.  

To illustrate this anomaly, we recall to the reader that the difference between the V4$_1$ and the Direct Fit (DF$_1$) models was that in the V4 model, the Regge powers $b_1$ and $b_2$ were fixed at 1/2, whereas they were allowed to vary in the Direct Fit model, {\em raising} the probability from $P_{\rm V4}=0.616$ to $P_{\rm V6}=0.721$, as expected---the {\em exact opposite} of the V4 to V5 effect, where the logarithmic power $\gamma$ was varied from 2.  Clearly, we question their minimization program, or their use of it.     
 
Thus, we conclude that there is no statistical evidence given in FMS that supports the conclusion that $\gamma > 2$, and thus, no basis for concluding that the Froissart bound is violated. 

{\em Prediction of the 7 TeV total $pp$ cross section}---Since Table 1 of FMS does not contain the Totem point, its data  could have been used to make  predictions of the total cross section at 7 TeV to compare with the Totem value, since it uses  only lower energy data, although the  FMS paper has not explicitly made any predictions.  However, their plots of their results in Tables 1 and 3 allow us to do so, visually. Inspection of the curve for the V4 model in Fig. 5 of FMS (the dash-dotted curve) shows that the central value of the 7 TeV  prediction (from the data using {\em only} the lower energies, labeled $\sqrt s_{\rm max}$ = 1.8 TeV) goes slightly inside the lower error bar of the plotted Totem result. To obtain their  7 TeV cross section prediction, together with its error, from their V4 model of Table 3 using their parameters in a standard error evaluation, we find $\sigma_{\rm tot}=96.2\pm 4.5$ mb, which is in excellent agreement with the experimental value of $98.1\pm 2.3$ found by Totem.      We see  indeed that the Totem value is well predicted using $\gamma=2$, with the FMS numbers that they themselves calculated. 

Similar conclusions may be drawn from the Direct Fit model of Table 1. Using the Direct Fit model parameters of  FSM's Table 1, we numerically calculate that the DF model predicts $95.4\pm 3.7$ mb when {\em only} the $1\sigma$  diagonal error due to the coefficient $\beta$ is taken into account ; it becomes $95.4\pm 8.8$ when  their diagonal error due to $s_h$ is included), so that their DF model prediction is also in good agreement with the Totem cross section, $98.1\pm 2.3$ mb.   

As in the preceding Section, we find that the FMS $\gamma=2$ fits at low energy, after allowing for errors in the predictions due to the statistical  errors in the fitting parameters, successfully  predict the Totem total cross section at 7 TeV, thus negating the necessity for considering a violation of the Froissart bound.  In simpler words, the FMS fits are consistent with a saturated Froissart bound when the Totem point is included.   

{\em Additional experimental evidence for a saturated Froissart bound}---Block and Halzen \cite{blackdisk2} (BH) have recently shown that the proton asymptotically becomes a black disk as $s\rightarrow\infty$, using analyticity constraints that anchor their fit at low energy to predict  ultra-high energy experimental data at the LHC at 7 TeV, as well as the Pierre Auger Observatory \cite{auger} results at 57 TeV. The BH model parameterizes the even and odd (under crossing) total cross sections and $\rho$-values and fits  4 experimental quantities, the 2 total cross sections $\sigma_{\bar pp}(\nu), \sigma_{p p}(\nu)$ and the 2 $\rho$-values, $\rho_{\bar pp}(\nu)$ and $\rho_{p p}(\nu)$, to the high energy analytic (complex) amplitude parameterizations \cite{bc}, which guarantee analyticity and are much simpler to compute than derivative dispersion relations. Using $\nu$ as the laboratory energy and denoting $m$ as the proton mass, they find that
\ba
\sigma^0(\nu)&\equiv&\beta_{\cal P'}\left(\frac{\nu}{m}\right)^{\mu -1}+c_0+c_1\ln\left(\frac{\nu}{m}\right)\nonumber\\
&&+c_2\ln^2\left(\frac{\nu}{m}\right),\label{sig0pp}\\
\sigma^\pm(\nu)&=&\sigma^0(\nu)\pm\  \delta\left({\nu\over m}\right)^{\alpha -1},\label{sigmapmpp}\\
\rho^\pm(\nu)&=&{1\over\sigma^\pm(\nu)}\left\{\frac{\pi}{2}c_1+c_2\pi \ln\left(\frac{\nu}{m}\right)\nonumber\right.\\
&&-\beta_{\cal P'}\cot({\pi\mu\over 2})\left(\frac{\nu}{m}\right)^{\mu -1}
\left.+\frac{4\pi}{\nu}f_+(0)\right.\nonumber\\
&&\left.\pm \delta\tan({\pi\alpha\over 2})\left({\nu\over m}\right)^{\alpha -1} \right\}\label{rhopmpp},
\ea
using the upper sign  for $pp$ and the lower sign  for  $\bar pp$; $\delta,\alpha,\beta_{\cal P'},\mu,c_0,c_1,c_2$ and  $f_+(0)$ are real constants, with $f_+(0)$ being the singly subtracted dispersion relation constant.  $\sigma^0(\nu)$, the even (under crossing) amplitude cross section of \eq{sig0pp} effectively becomes
\ba
\sigma^0(s)&=&c_0+c_1\ln\left(\frac{s}{2m^2}\right)
+c_2\ln^2\left(\frac{s}{2m^2}\right),\label{sig0pp1}
\ea 
so that asymptotically, $\sigma^0\rightarrow \ln^2 s$. 
Thus, the high energy functional form used by FMS, \eq{sigmaHE}, when $\gamma=2$, is completely equivalent to the functional form of the above \eq{sig0pp1}. In other word, BH replace the 3 FMS parameters $\alpha,\ \beta$ and $s_h$ by the 3 linear parameters $c_0,\ c_1$ and $c_2$. Thus, the BH fit uses approximately the same physical assumptions as the V4 model in Table 3 of FMS, fitting data for cms energies  $\sqrt s\le 1800$ GeV. 

\begin{figure}[h]%Fig. 3
\begin{center}
\mbox{\epsfig{file=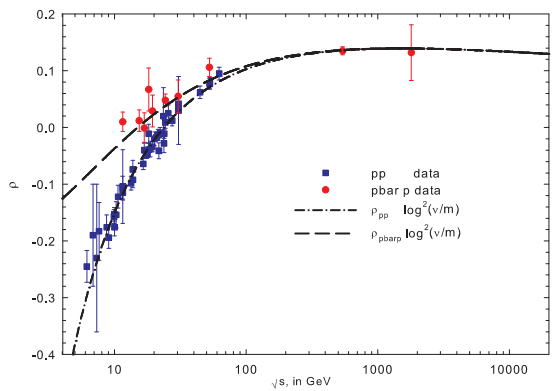
,width=3in%
,bbllx=42pt,bblly=291pt,bburx=585pt,bbury=660pt,clip=%
}}
\end{center}
\caption[ ]{\protect {Froissart-bounded analytic amplitude fits to $\rho$, the ratio of the real to the imaginary portion of the forward scattering amplitude, vs. $\sqrt s$, the cms energy in GeV, taken from BH \cite{blockhalzen2}. 
  The $\bar pp$ data used in the fit  are the (red) circles and the $pp$ data are the (blue) squares. }}
\label{fig:rho}
\end{figure}

The results \cite{blockhalzen2} of a {\em global} fit to both $\bar p p$ and $pp$ $\rho$-values and total cross sections in the energy range $6\le\sqrt s\le 1800$ GeV, with 187 datum points and {\em only} 5 free parameters, $\delta,\alpha,c_1,c_2$ and $f_+(0)$,  is shown in Fig. \ref{fig:rho}. 
%%%%%%%%%%%%%%%%%%%%  
As seen from \eq{rhopmpp},  $\rho\rightarrow 0$ as $s\rightarrow\infty$, which is a requirement for a black disk at infinity. However,  the tiny change in $\rho$ from 0.135 at 1800 GeV to 0.132 at 14000 GeV implies that we are nowhere near asymptopia, where $\rho=0$.

\begin{figure}[h,t,b] %Fig.1
\begin{center}
\mbox{\epsfig{file=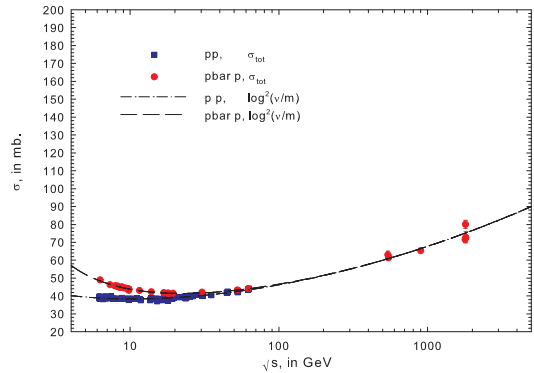
,width=3in%
,bbllx=90pt,bblly=425pt,bburx=555pt,bbury=730pt,clip=%
}}
\end{center}
\caption[]{
 Froissart-bounded analytic amplitude fits to the total cross section, $\sigtot$, in mb, for $\bar pp$ (dashed curve)  and $pp$ (dot-dashed curve)  from \eq{sigmapmpp}, vs. $\sqrt s$, the cms energy in GeV, taken from BH  \cite{blockhalzen2}. The $\bar pp$ data used in the fit  are the (red) circles and the $pp$ data are the (blue) squares.  The fitted data were anchored by values of $\sigtot^{\bar pp}$ and $\sigtot^{pp}$, together with the energy derivatives  ${d\sigtot^{\bar pp}/ d\nu}$ and ${d\sigtot^{pp}/ d\nu}$ at 6 GeV using FESR, as described in Ref. \cite{blockhalzen2}. We note that their ultra-high energy total cross section predictions that are made from their analytic amplitude  fit use {\em only}  total cross section  data that are in the lower energy range  $6\le \sqrt s \le 1800$ GeV.
\label{fig:ppfit}
}
\end{figure}

The fits for the $pp$ and $\bar pp$ total cross sections, that use only 4 of the {\em same} parameters that were used for the $\rho$-value fit of Fig. \ref{fig:rho}, are shown in Fig. \ref{fig:ppfit}. 
%%%%%%%%%%%%%%%%%%%%%%%%%%%
%%%%%%%%%%%%%%%
The dominant $\ln^2 s$ term in the total cross section $\sigma^0$ (see \eq{sig0pp1}) saturates the Froissart bound \cite{froissart}; thus it controls the ultra-high energy behavior of the total cross sections.

\begin{figure}[h]%
\begin{center}
\mbox{\epsfig{file=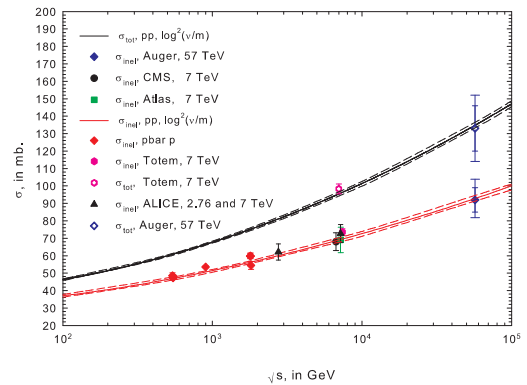
,width=3in%
,bbllx=33pt,bblly=390pt,bburx=510pt,bbury=715
pt,clip=%
}}
\end{center}
\caption[]{\protect
{ Predictions for $\sigma_{\rm tot}$ and $\sigma_{\rm inel}$, in mb, vs. $\sqrt s$, in GeV,  for $\bar pp$ and $pp$, taken from Block and Halzen \cite{blackdisk2}. For $\sigma_{\rm tot}$, they have compared their predictions with recent $pp$ TOTEM data at 7 TeV and Auger data at 57 TeV, while for $\sigma_{\rm inel}$, they have compared their results with  2.76 TeV $pp$ data from ALICE, 7 TeV $pp$ data from ALICE, ATLAS, CMS and Totem, as well as with the 57 TeV  $pp$ inelastic cross section. The upper solid (black) curve is the central-value prediction for $\sigma_{\rm tot}$ and the lower solid (red) curve is the central-value prediction for $\sigma_{\rm inel}$.  The dotted curves are the errors ($\pm 1 \sigma$) in their predictions, due to the correlated errors of the fitting parameters.     
}
}
\label{fig:pppredictions}
\end{figure}

A very important role in fixing the overall fit is played by Finite Energy Sum Rules (FESR) \cite{blockanalyticity,physicreports} anchoring the fit at the low energy end, taken to be $\sqrt s=6$ GeV. The FESR allow BH \cite{blockhalzen2} to { fix the fit} to both $\sigtot^{\bar pp}$ and $\sigtot^{pp}$, together with their two energy derivatives  ${d\sigtot^{\bar pp}/ d\nu}$ and ${d\sigtot^{pp}/ d\nu}$, at the low end, $\sqrt s= 6$ GeV, by  using the { many precise low energy total cross section measurements}  between $\sqrt s$ of 4 and 6 GeV.  These FESR constraints then can be used to evaluate directly   the values of  $c_0$ and $\beta_{\cal P'}$, two of the four parameters needed to determine $\sigma^0$, the even high energy total cross section of \eq{sig0pp}.  These fixed values of  $c_0$ and $\beta_{\cal P'}$, together with with the 2 {\em  globally fitted} values of $c_1$ and $c_2$  required for $\sigma^0(\nu)$ (obtained from fitting {\em simultaneously}  the high energy total cross section and $\rho$ measurements in the energy region $6\le \sqrt s \le 1800$ GeV), are listed in Table \ref{tab:sigma0}. We remind the reader that only data in the energy region $6\le \sqrt s\le 1800$ GeV are used in this global fit (together with the prolific and accurate 4 to 6 GeV total cross section  data used for the  6 GeV low energy `anchor points'). We note that  $c_2$, the coefficient of $\ln^2(s)$, is well-determined, having a statistical accuracy of $\sim 2\%$. Thus, we see from Fig. \ref{fig:ppfit} that the  experimental data show that a saturated Froissart bound model is accurately satisfied for $\sigtot$, the total cross sections for both $\bar p p$ and for $pp$ in the energy interval $6\le \sqrt s\le 1800$ GeV; this accuracy of prediction mainly results from the use of the FESR constraints on the high energy  analytic amplitude fit \cite{blockanalyticity,physicreports}.
Ultra-high energy  total cross sections, for which there are no distinction between $\bar p p$ and $pp$ interactions---both being given by $\sigma^0$---are now accurately  predicted. For example, we obtain values for the total $pp$ cross section of $\sigma^0=95.4\pm 1.1$ mb at 7 TeV \cite{7tev} and $134.8\pm 1.5$ mb at 57 TeV \cite{cr}, where the $\pm 1\sigma$ errors are calculated from the (correlated) errors of their fit parameters $c_1$ and $c_2$. The (black) upper curves in Fig. \ref{fig:pppredictions} are plots of these BH \cite{blackdisk2}  predictions for the total cross section $\sigma^0$ vs. $\sqrt s$, with the solid curve being the central value and the dashed curves the $\pm 1\sigma$ error curves. 

%%%%%%%%%%%%%%%%%%%%%%%%%%%%%%%%%%%%%%%%%%%%%%%%
\begin{table}[h,t]                   % Use "table" environment, but also
				 % use  "tabular" environment below.
%
\def\arraystretch{1.5}            % Make the space between rows in the Table,
				  % 1.5 x bigger than the default spacing.

\begin{center}
\caption[]{Values  of the parameters, in mb, needed for the even amplitude total cross section, $\sigma^0(\nu)$ of \eq{sig0pp}, taken from Ref. 
\cite{blockhalzen2}.
\label{tab:sigma0}
}
\vspace{.2in}
\begin{tabular}[b]{||c||c||}
\hline\hline
$c_0$=$37.32$ mb,&$\beta_{\cal P'}$=$37.10$ mb\\
\hline
$c_1$=$-1.440\pm 0.070$ mb,&$c_2$=$0.2817\pm 0.0064$ mb,\\
\hline\hline
\end{tabular}
     %\vspace{1in} \\
\end{center}
\end{table}
\def\arraystretch{1}  %Restore the default row spacing in the Table.
%%%%%%%%%%%%%%%%%%%%%%%%%%%%%%%%%%%

The inelastic cross section, $\sigma_{\rm inel}^0$, is determined by numerically multiplying the ratio of the inelastic to total cross section with the fitted total cross section $\sigma^0$.  The ratio of inelastic to total cross section was  determined from an eikonal model, called the `Aspen' model; for details see Ref. \cite{aspen} and Ref. \cite{blackdisk}. 

Block and Halzen  \cite {blackdisk} found that 
\ba
\sigma_{\rm inel}^0(\nu)&=& 62.59\left(\frac{\nu}{m}\right)^{-0.5}+24.09+0.1604 \ln\left(\frac{\nu}{m}\right)\nonumber\\
&&+ 0.1433 \ln^2\left(\frac{\nu}{m}\right) \ {\rm mb},
\label{finalinelastic}
\ea 
valid in the energy domain, $\sqrt s \ge 100$ GeV.

The lower (red) plots of Fig. \ref{fig:pppredictions} are the BH  \cite{blackdisk2} predictions (\eq{finalinelastic} for  high energy  {\em inelastic} cross sections $\sigma_{\rm inel}$, as a function of  $\sqrt s$. 
 %%%%%%%%%%%%%%%%%%%%  
%%%%%%%%%%%%%%%%%%%%  
The solid (red) plot is the central value and the error bands corresponding to $\pm 1 \sigma$ are the lower dashed (red) curves. All of the existing inelastic cross section measurements for $\bar p p$, as well as the eight new ultra-high $pp$ measurements, are shown. The agreement with experiment is excellent over the entire energy scale.  In particular, the agreement with the new highest energy (57 TeV)  experimental measurements of both $\sigma_{\rm tot}$ and $\sigma_{\rm inel}$ is striking. Since {\em none} of the experimental datum points in Fig. \ref{fig:pppredictions} were used in making these predictions, it is clear that the $\ln^2 s$ predictions for $\sigma_{\rm inel}$ and $\sigma_{\rm tot}$ are strongly supported by the existing ultra-high energy measurements. 

Finally,  BH \cite{blackdisk}  determined the ratio of $\sigma_{\rm inel}(s)/\sigma_{\rm tot}(s)$ as $s\rightarrow \infty$,  given by the ratio of  the  $\ln^2 s$ coefficients  in $\sigma^0_{\rm inel}$ and $\sigma^0$, respectively, i.e.,
\ba  
{\sigma_{\rm inel}\over \sigma_{\rm tot}}\rightarrow {c_2^{\rm inel}\over c_2}={0.1433\over 0.2817}=0.509\pm 0.021,\quad {\rm as\ } s\rightarrow \infty,
\ea
that is well within error of the expected value of $\frac{1}{2}$ that is appropriate for a black disk at infinity.

{\em Conclusions}---We find that:
\begin{enumerate}
  \item there is no sound statistical evidence put forth by FMS  \cite {FMS} for their conclusion that the Froissart bound is exceeded. Their models with $\ln^2 s$ that only use lower energy cross sections actually {\em predict} the Totem total $pp$ cross section reasonably well.
  \item 
 {\em both} the measured total $pp$ cross sections and the inelastic cross sections are fit up to $\sqrt s=57$ TeV by a saturated Froissart-bounded $\ln^2s$ behavior that is associated with a black disk.
\item the forward scattering amplitude is pure imaginary as $s\rightarrow \infty$, as is  required for a black disk.
\item the ratio of ${\sigma_{\rm inel}/ \sigma_{\rm tot}}\rightarrow 0.509\pm 0.021,\quad {\rm as\ } s\rightarrow \infty$, compatible with the black disk value of 0.5.
\end{enumerate}
Thus, we conclude that existing experimental evidence strongly supports the conclusion that the proton becomes a black disk at infinity, whose total cross section goes as $\ln^2 s$ as $s\rightarrow \infty$.

{\em Acknowledgments}---In part,  F. H. is supported by the National Science Foundation  Grant No. OPP-0236449, by the DOE  grant DE-FG02-95ER40896 and  by the University of Wisconsin Alumni Research Foundation. M. M. B. thanks  the Aspen Center for Physics, supported in part by NSF Grant No. 1066293,  for its hospitality during this work.
%%%%%%%%%%%%%%%%%%%%%%%%%%%%%%%%%%

%
\end{document}